\documentclass[12pt,a4paper]{article}
\usepackage{graphicx}
\usepackage{amssymb,amsfonts,amsmath}
\usepackage{color}
\usepackage{cite}
\begin{document}

\newcommand{\EQ}{Eq.~}
\newcommand{\EQS}{Eqs.~}
\newcommand{\FIG}{Fig.~}
\newcommand{\FIGS}{Figs.~}
\newcommand{\TAB}{Tab.~}
\newcommand{\TABS}{Tabs.~}
\newcommand{\SEC}{Sec.~}
\newcommand{\SECS}{Secs.~}

\newcommand{\comr}[1]{\textcolor{red}{#1}}
\newcommand{\comb}[1]{\textcolor{blue}{#1}}

\title{Impact of hierarchical modular structure on ranking
of individual nodes in directed networks}
\author{Naoki Masuda${}^{1,2*}$, Yoji Kawamura${}^{3}$,
 and Hiroshi Kori${}^{4,2}$\\
\ \\
\ \\
${}^{1}$ 
Graduate School of Information Science and Technology,\\
The University of Tokyo,\\
7-3-1 Hongo, Bunkyo, Tokyo 113-8656, Japan
\ \\
${}^2$
PRESTO, Japan Science and Technology Agency,\\
4-1-8 Honcho, Kawaguchi, Saitama 332-0012, Japan
\ \\
${}^3$ Institute for Research on Earth Evolution,\\
Japan Agency for Marine-Earth Science and Technology,\\
3173-25 Showa-machi, Kanazawa-ku, Yokohama, Kanagawa 236-0001, Japan
\\
${}^4$ Division of Advanced Sciences, Ochadai Academic Production,\\
Ochanomizu University,\\
2-1-1, Ohtsuka, Bunkyo-ku, Tokyo 112-8610, Japan\\ 
\ \\
$^*$ Author for correspondence (masuda@mist.i.u-tokyo.ac.jp)}

\setlength{\baselineskip}{0.77cm}
\maketitle

\newpage

\begin{abstract}
\setlength{\baselineskip}{0.77cm} 
Many systems, ranging from biological and engineering systems to social systems, can be modeled as directed networks, with links representing directed interaction between two nodes. To assess the importance of a node in a directed network, various centrality measures based on different criteria have been proposed. However, calculating the centrality of a node is often difficult because of the overwhelming size of the network or the incomplete information about the network. Thus, developing an approximation method for estimating centrality measures is needed. In this study, we focus on modular networks; many real-world networks are composed of modules, where connection is dense within a module and sparse across different modules. We show that ranking-type centrality measures including the PageRank can be efficiently estimated once the modular structure of a network is extracted. We develop an analytical method to evaluate the centrality of nodes by combining the local property (\textit{i.e.}, indegree and outdegree of nodes) and the global property (\textit{i.e.}, centrality of modules). The proposed method is corroborated with real data. Our results provide a linkage between the ranking-type centrality values of modules and those of individual nodes. They also reveal the hierarchical structure of networks in the sense of subordination (not nestedness) laid out by connectivity among modules of different relative importance. The present study raises a novel motive of identifying modules in networks.
\end{abstract}

\newpage

\section{Introduction}\label{sec:introduction}

A variety of systems of
interacting elements can be represented as
networks. A network is a collection of nodes and links; a
link connects a pair of nodes. Generally speaking, 
some nodes play central functions, such as
binding different parts of the network together and controlling
dynamics in the network. To identify important nodes in a
network, various centrality
measures based on different criteria have been proposed
\cite{Wasserman94,Newman03siam,Boccaletti06}.

Links of many 
real networks such as the World Wide Web (WWW), food webs, neural
networks, protein interaction networks, and many social networks are
directed or asymmetrically weighted. In contrast to the case of
undirected networks, a link in directed networks indicates an
asymmetrical relationship between two nodes, for example, the control of
the source node of a link over the target node. The
direction of a link indicates the relative importance of the two nodes.
Central nodes in a network
in this sense would be, for example, executive personnels in
an organizational network and top predators in a food web. 
Generally, more (less) central nodes are located at
an upper level (a lower level) in the hierarchy of the network,
where hierarchy
refers to the distinction between upper and lower levels 
in terms of the centrality value as
relevant in, for example, biological
\cite{Garlaschelli03nat,Lagomarsino07pnas} and social
\cite{Castellano09} systems.
This type of
centrality measure is necessarily specialized for directed networks
and includes the popularity or prestige measures
for social networks \cite{Wasserman94
}, ranking systems for webpages such as the PageRank \cite{Brin98,Berkhin05}
and HITS \cite{Kleinberg99,Kleinberg01}, adaptations of the PageRank 
to citation networks
of academic papers \cite{Palacios04,ChenRedner07
}
and  journals \cite{Palacios04,Pinski76,
Davis08,Fersht09}, and
ranking systems of sports teams \cite{Park05jsm}.
We call them ranking-type centrality measures.

Under practical restrictions such as overwhelming network size or
incomplete information about the network, it is often difficult to
exactly obtain ranking-type centrality values of nodes.
%
%
In such situations, the simplest
approximators are perhaps those based on the degree of nodes
(\textit{i.e.}, the number of links owned by a node).  For example,
the indegree of a node
can be an accurate approximator of the PageRank of websites
\cite{Fortunato06pnas
} and ranks of academic journals \cite{Davis08,Fersht09}.
However, such local approximations often fail
\cite{Donato04,Restrepo06prl,Masuda09njp}, implying a significant effect
of the global structure of networks.
%

A ubiquitous global structure of networks that adversely affects
local approximations is the modular structure.
Both in undirected \cite{Newman04epjb,Palla05nat,
Palla07nat,Fortunato09} and directed
\cite{Wasserman94,Fortunato09,Palla07newjp,Leicht08prl,Rosvall08pnas
} networks, nodes are often classified into modules (also called
communities) such that the nodes are densely connected within a module
and sparsely connected across different modules.  In modular networks,
some modules may be central in a coarse-grained network, where each
module is regarded as a supernode \cite{Everett99}.  However,
relationships between the centrality of individual nodes and that of
modules are not well understood.  Using these relationships, we will
be able to assess centralities of individual nodes only on the
basis of coarse-grained information about the organization of modules
or under limited computational resources.

In this study, we analyze the ranking-type centrality measures for
directed modular networks.  We are concerned with the modular
structure of the network in the meaning of partitioning of the network
into parts, and not the overlapping community structure
\cite{Palla05nat,Palla07nat,Palla07newjp}.  We determine the
centrality of modules, which reflects the hierarchical structure
of the networks in the sense of subordination
\cite{Garlaschelli03nat,Lagomarsino07pnas,Castellano09}, not
nestedness \cite{Ravasz02,Guimera05nat,Sales07,Clauset08nat}. Then,
we show that module membership is a chief determinant of the
centrality of individual nodes.  A node tends to be central when it
belongs to a high-rank module and it is locally central by, for
example, having a large degree. To clarify these points, we
analytically evaluate centrality in modular networks.  On the basis of
the matrix tree theorem, the centrality value of a node is derived
from the number of spanning trees rooted at the node. We use this
relationship to develop an approximation scheme for the ranking-type
centrality values of nodes in modular networks.  The approximated
value turns out to be a combination of local and global effects,
\textit{i.e.}, the degree of nodes and the centrality of modules.
For analytical tractability, we formulate our theory using the
ranking-type centrality measure called the influence, but the results
are also applicable to the PageRank. We corroborate the
effectiveness of the proposed scheme using the \textit{Caenorhabditis elegans}
neural network, an email social network, and the WWW.


\section{Ranking-type Centrality Measures}

We consider a directed and weighted network of $N$
nodes denoted by $G=\{V,E\}$. A set of nodes is denoted by
$V=\{1,\ldots, N\}$, and 
$E$ is a set of directed links, \textit{i.e.},
node $i$ sends a directed link
to node $j$ with weight $w_{ij}$
if and only if $(i,j)\in E$. The weight represents the amplitude
of the direct influence of node $i$ on node $j$.
We set $w_{ij}=0$ when
$(i,j)\notin E$. 

Depending on applications, 
different centrality measures can be used
to rank the nodes in a network.
We analyze the effect of the modular
structure on ranking of nodes using a centrality measure called
\textit{influence} because it facilitates theoretical analysis.
The existence of a one-to-one mapping
from the influence to
the PageRank \cite{Brin98,Berkhin05,Fortunato06pnas,Donato04} 
and to variations of the PageRank 
used for ranking academic journals and articles
\cite{Palacios04,Pinski76,Davis08,Fersht09}, which we will explain
in this section, 
enables us to adapt our results to the
case of such ranking-type centrality measures.  To show that our
results are not specific to the proposed measure, we study the
influence and the PageRank simultaneously.

We define the influence of node $i$, denoted by $v_i$, by the solution
of the following set of $N$ linear equations:
\begin{equation}
v_i = 
\frac{\sum_{j=1}^N w_{ij}v_j}{k_i^{\rm in}},\quad (1\le i\le N),
\label{eq:influence}
\end{equation}
where $k_i^{\rm in}\equiv \sum_{j^{\prime}=1}^N w_{j^{\prime} i}$
is the indegree of node $i$, and
$\sum_{i=1}^N v_i=1$ provides the normalization.  $v_i$ is
large if (i) node $i$ directly affects many nodes (\textit{i.e.}, many
terms probably with a large $w_{ij}$ on the RHS of
\EQ\eqref{eq:influence}), (ii) the nodes that receive directed links from
node $i$ are influential (\textit{i.e.}, large $v_j$ on the
RHS), and (iii) node $i$ has a small indegree.

Equation~\eqref{eq:influence} is the definition for strongly connected
networks; $G$ is defined to be strongly connected if there is a path,
\textit{i.e.}, a sequence of directed links,
from any node $i$ to any node $j$.
If $G$ is not strongly connected,
there is no path from a certain node $i$ to a certain node $j$.  Then,
node $i$ cannot influence node $j$ even indirectly, and the problem of
determining the influence of nodes is decomposed into that for
each strongly connected component.  Therefore, we assume that $G$ is
strongly connected.

The influence $v_i$ represents the importance of nodes in
different types of dynamics on networks (see Appendix~A
for details). Firstly, $v_i$ is equal to the fixation
probability of a new opinion introduced at node $i$
in a voter-type interacting particle 
system \cite{Masuda09njp}.
Secondly, if all links are reversed such that
a random walker visits influential nodes with high probabilities,
$v_i$ is the stationary density of the continuous-time simple
random walk.
Thirdly, $v_i$ is the so-called reproductive value
used in population ecology \cite{Taylor90,Taylor96}. Fourthly,
$v_i$ is the contribution of an opinion at node $i$ to the opinion of
the entire population in the consensus in the
continuous-time version of the 
DeGroot model
\cite{Degroot74,Jacksonbook,Olfati07}.
Fifthly, $v_i$ is equal to the amplitude of the
collective response in the synchronized dynamics when an input is
given to node $i$ \cite{Kori08}.

The influence can be mapped to the PageRank.
The PageRank, denoted by $R_i$ for node $i$, is defined
self-consistently by
\begin{equation}
R_i=
\frac{q}{N}+(1-q)\sum_{j=1}^N  \frac{w_{ji}}{k_j^{\rm out}} R_j
+\delta_{k_i^{\rm out},0}(1-q)R_i,\quad (1\le i\le N)
\label{eq:page}
\end{equation}
where $k_i^{\rm out}\equiv \sum_{j^{\prime}=1}^N w_{ij^{\prime}}$ 
is the outdegree of node $i$, 
$\delta_{i,j}=1$ if $i=j$, and $\delta_{i,j}=0$ if
$i\neq j$.
The second term on the RHS of \EQ\eqref{eq:page} is present
only when $k_j^{\rm out}>0$.
Note that the direction of the link 
in the PageRank has the meaning opposite
to that in the influence; $R_i$ of a webpage is incremented
by an incoming link (hyperlink),
whereas $v_i$ is incremented by an outgoing link.
The introduction of $q>0$ homogenizes $R_i$ and is necessary
for the PageRank to be defined for directed networks that are
not strongly connected, such as real web graphs.
The normalization is given by $\sum_{i=1}^N R_i=1$.
$R_i$ is regarded as the stationary
density of the discrete-time simple random walk on the network
\cite{Brin98,Berkhin05,Fortunato06pnas}, 
where $q$ is the probability of a jump to a randomly selected node.

%
An essential difference between the two measures lies in
normalization. In the influence, the total credit that node $j$ gives
its neighbors is equal to $\sum_{i=1}^N w_{ij}v_j=k_j^{\rm
in}v_j$, while that in the PageRank is equal to
$\sum_{i=1}^N \left(w_{ji}/k_j^{\rm out}\right)R_j=R_j$.
In the PageRank, the multiplicative factor of the total
credit that node $j$ gives other nodes is set to $\sum_{i=1}^N
w_{ji}/k_j^{\rm out}=1$ to prevent nodes with many outgoing links from
biasing ranks of nodes. In the ranking of webpages, creation of a
webpage $i$ with many hyperlinks does not indicate that node $i$ gives
a large amount of credit to recipients of a link.
%
%
Each neighbor of node $i$ receives the credit $R_i/k_i^{\rm
out}$ from node $i$.  We should refer to the PageRank when nodes can
select the number of recipients of credit (\textit{e.g.}, the
WWW and citation-based ranking of academic papers and journals).  We
should use the influence when the importance
of all links is proportional to their weights
(\textit{e.g.}, opinion formation and
synchronization mentioned above).


The PageRank is equal to the influence in a network modified
from the original network $G$ (see Appendix~B for derivation).
In particular, the PageRank in $G$ for $q=0$ is
given by
\begin{equation}
R_i= k_i^{\rm out}v_i\left(G^{\rm rev}\right),
\label{eq:R vs v}
\end{equation}
where $v_i\left(G^{\rm rev}\right)$ is 
the influence of node $i$ for the network $G^{\rm rev}$, which is 
obtained by reversing all links of $G$.
We use this relation
to extend our results derived for
the influence to the case of the PageRank. 

The influence has a nontrivial sense only in
directed networks because
$w_{ij}=w_{ji}$ in \EQ\eqref{eq:influence}
leads to $v_i=1/N$ \cite{Masuda09njp,Antal06prl,Sood08pre}.
Furthermore, any network
with $k_i^{\rm in}=k_i^{\rm out}$ ($1\le i\le N$)
results in $v_i=1/N$. Therefore, from
\EQ\eqref{eq:R vs v},
$R_i= k_i^{\rm in}/\left(\left<k\right>N\right)$
for such a network, where
$\left<k\right>\equiv \sum_{i=1}^N
k_i^{\rm in}/N=\sum_{i=1}^N k_i^{\rm out}/N$ is
the mean degree. In this case, 
$v_i$ and $R_i$ are not affected by the global structure of the
network. 

In directed or asymmetrically weighted networks,
$v_i$ and $R_i$ are heterogeneous in general. 
The mean-field approximation (MA)
is the simplest ansatz based on the local property of a node.
By using
$\sum_{j=1}^N w_{ij} v_j
\approx \sum_{j=1}^N w_{ij} \bar{v} = k_i^{\rm out} \bar{v}$,
where $\bar{v}\equiv \sum^N_{i=1}v_i/N=1/N$,
we obtain $v_i \propto k_{i}^{\rm out}/k_{i}^{\rm in}$.
Combination of this and \EQS\eqref{eq:R vs v}
yields the MA for the PageRank:
$v_i\approx k_i^{\rm in}/\left(\left<k\right>N\right)$.


We can calculate $v_i$ by enumerating spanning trees.
To show this, note that \EQ\eqref{eq:influence} implies that
$v_i$ is the left eigenvector with eigenvalue zero
of the Laplacian matrix 
defined by
$L_{ii}=\sum_{j^{\prime}=1}^N w_{j^{\prime}i}$ 
and $L_{ij}=-w_{ji}$ ($i\neq j$), \textit{i.e.},
\begin{equation}
\sum^N_{i=1}v_iL_{ij}=0,\quad (1\le j\le N).
\end{equation}
The $(i,j)$ cofactor of $L$ is defined by
\begin{equation}
{\rm Co}\left( i, j \right)
\equiv (-1)^{i+j} \det{L}\left( i, j \right),
\label{eq:adjoint}
\end{equation}
where $L(i,j)$ is an $(N-1)\times (N-1)$ matrix obtained by deleting
the $i$-th row and the $j$-th column of $L$.
Because $\sum_{j=1}^N L_{ij} = 0$, ($1\le i\le N$),
${\rm Co}\left( i, j \right)$ does not depend on $j$.
Using \EQ\eqref{eq:adjoint} and the fact that
$L$ is degenerate, we obtain
\begin{align}
  \sum_{i=1}^N {\rm Co}(i,i) L_{ij}
  &= \sum_{i=1}^N {\rm Co}(i,j)L_{ij}\nonumber\\
  &= \det L = 0,\quad (1\le j\le N).
\end{align}
Therefore, $\left({\rm Co}\left( 1, 1 \right),
\ldots, {\rm Co}\left( N, N \right)\right)$ is the left eigenvector of
$L$ with eigenvalue zero, which yields
\begin{equation}
v_i\propto {\rm Co}\left( i,i\right) 
=\det L\left( i,i\right).
\end{equation}
From the matrix tree theorem \cite{ref:biggs,ref:agaev},
$\det L\left( i,i\right)$
is equal to the sum of the weight of 
all possible directed spanning trees rooted at node $i$. The
weight of a spanning tree is equal to the product of 
the weight of $N-1$ links forming the spanning tree.

\section{Centrality in Modular Directed Networks}

\begin{figure}
\begin{center}
\includegraphics[width=8cm]{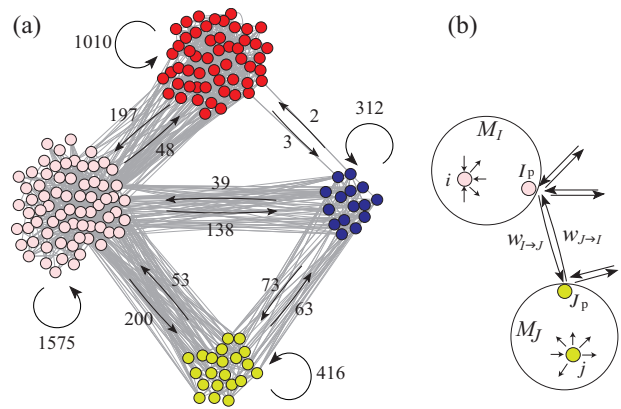}
\caption{(a) A part of \textit{C.~elegans} neural network
composed of the 4 largest modules.
The link weight is equal to the sum of the number of
chemical synapses and that of the gap junctions.
The original network has 274 nodes, 2959 links, and 13 modules, while
the depicted subnetwork has 159 nodes 1363 links.
The values indicate the summed link weights
from one module to another.
(b) Approximation of
intermodular connectivity by links between portal nodes.}
\label{fig:schem modular}
\end{center}
\end{figure}

Most directed networks in the real world are more
structured than those captured by the MA.
A ubiquitous global structure of networks is modular
structure.  Modular networks consist of several densely connected
subgraphs called modules (also called communities),
and modules are connected to each other by
relatively few links. As an example,
a subnetwork of the \textit{C.~elegans} neural network 
\cite{Chen06pnas,wormatlas} containing 4
modules is shown in \FIG\ref{fig:schem modular}(a).
Modular structure is common in both
undirected \cite{Newman04epjb,Palla05nat,Palla07nat,Fortunato09}
and directed
\cite{Fortunato09,Palla07newjp,Leicht08prl,Rosvall08pnas} networks.

Modular structure of directed networks often leads to hierarchical
structure. By hierarchy, we refer to the situation in
which modules are located at different levels
in terms of the value of the ranking-type centrality.
It is relatively easy
to traverse from a node in an upper level to one in a lower level
along directed links, but not vice versa.  The hierarchical structure
leads to the deviation of $v_i$ from the value obtained from
the MA.  

\begin{figure}
\begin{center}
\includegraphics[width=8cm]{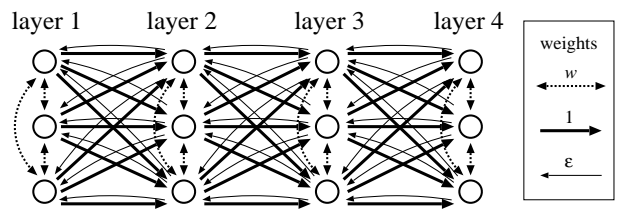}
\caption{Hierarchical multipartite network with $N=12$ and $P=4$.}
\label{fig:example hie}
\end{center}
\end{figure}

As an example,
consider the directed $P$-partite network shown in
\FIG\ref{fig:example hie}.  Layer $P^{\prime}$ ($1\le P^{\prime} \le P$)
contains $N/P$ nodes, where $N$ is divided by $P$. The nodes in the
same layer are connected bidirectionally with weight $w$.  Each node
in layer $P^{\prime}$ ($1\le P^{\prime}\le P-1$) sends directed links
to all nodes in layer $P^{\prime}+1$ with weight unity, and each node
in layer $P^{\prime}$ ($2\le P^{\prime}\le P$) sends directed links to
all the $N/P$ nodes in layer $P^{\prime}-1$ with weight
$\epsilon$. The following results do not change if two adjacent layers
are connected via just an asymmetrically weighted bridge,
as shown in \FIG\ref{fig:schem modular}(b).
%
%
Because of the symmetry, all nodes in layer $P^{\prime}$ have the same
influence $v_{P^{\prime}}$. From \EQ\eqref{eq:influence}, we
obtain
\begin{equation}
v_{P^{\prime}}
=\frac{\epsilon^{P^{\prime}-1}(1-\epsilon)P}{(1-\epsilon^P)N}.
\label{eq:vi chain special}
\end{equation}
When $\epsilon<1$, a node in a layer with small $P^{\prime}$ 
is more influential than a node in
layer with large $P^{\prime}$.
The MA yields
\begin{equation}
\frac{k_i^{\rm out}}{k_i^{\rm in}}=
\left\{\begin{array}{ll}
\epsilon^{-1}, & (\mbox{node } i\in \mbox{ layer } 1)\\
\epsilon, & (\mbox{node }i\in \mbox{ layer } P)\\
1, & (\mbox{otherwise})
\end{array}\right.
\end{equation} 
%
%
%
The actual $v_{P^{\prime}}$ decreases exponentially throughout the
hierarchy, whereas $k_i^{\rm out}/k_i^{\rm in}$ does not. We observe a
similar discrepancy in the case of the PageRank.

We develop an improved approximation for
the influence in modular networks by
combining the MA and the correction factor obtained from the global
modular structure of networks. Consider a
network of $m$ modules $M_I$ ($1\le I\le m$).
For mathematical tractability,
we assume that each module communicates with the other
modules via a single portal node $I_p\in M_I$, as illustrated in
\FIG\ref{fig:schem modular}(b); the network shown in 
\FIG\ref{fig:schem modular}(b) is an approximation of
that shown in \FIG\ref{fig:schem modular}(a).
We denote the weight of the link
$(I_p, J_p)$ by $w_{I\to J}$ ($I\neq J$).

We obtain $v_i$ in this modular network by enumerating spanning trees
rooted at node $i\in M_I$. Denote such a spanning tree by $T$.
The intersection of $T$
and $M_I$ is a spanning tree restricted to $M_I$ and rooted at
node $i$.  This restricted spanning tree reaches all nodes in $M_I$.
$T$ enters $M_J$ ($J\neq I$) via a directed path from node $I_p$ to
node $J_p$. This path
is provided by a spanning tree in the network of $m$ modules, where
each module is represented by a single node.
The other nodes in
$M_J$ are spanned by the intersection of $T$
and $M_J$, which forms a spanning
tree restricted to $M_J$ and rooted at node $J_p$. Therefore, $T$
is a concatenation of (i) an intramodular
spanning tree in $M_I$ and rooted at node $i$, (ii) $m-1$
intramodular spanning trees in $M_J$ and rooted at
node $J_p$ ($J\neq I$), and (iii)
a spanning tree in the network of $m$ modules
rooted at node $I_p$. Let ${\cal N}_{\ell}(M_I)$
($\ell$ for local) denote the number of spanning
trees in $M_I$ with an arbitrary root, and
${\cal N}_g$ ($g$ for global) denote the number of spanning trees in a
network of $m$ modules with an arbitrary root.
Then, the number of spanning trees in $G$ rooted at node $i$ is equal to
\begin{equation}
{\cal N}_{\ell}(M_I)v_i^{\ell}
\left[\prod^m_{J=1,J\neq I} \left({\cal N}_{\ell}\left(M_J\right)
v_{J_p}^{\ell}\right)\right]
{\cal N}_g v_{M_I}^g,
\label{eq:enumeration}
\end{equation}
where $v_i^{\ell}$ is the influence of
node $i\in M_I$ within $M_I$ and
$v_{M_I}^g$ is the 
influence of $M_I$ in the network of $m$ modules.
The first, second, and third factors in \EQ\eqref{eq:enumeration}
corresponds to the numbers of spanning trees of types (i), (ii), and
(iii), respectively. Therefore, we obtain
\begin{equation}
v_i \propto v_i^{\ell} \left(\prod^m_{J=1,J\neq I}
v_{J_p}^{\ell}\right)v_{M_I}^g.
\label{eq:modular spanning}
\end{equation}

For nodes $i, i^{\prime}\in M_I$,
\EQ\eqref{eq:modular spanning} yields
$v_i/v_{i^{\prime}}=v_i^{\ell}/v_{i^{\prime}}^{\ell}$;
the relative influence
of nodes in the same module is equal to
their relative influence within the module.
For nodes in different modules, \textit{i.e.},
node $i$ in module $M_I$ and node $j$ in module $M_J$ ($I\neq J$),
\EQ\eqref{eq:modular spanning} leads to
\begin{equation}
\frac{v_i}{v_j}
=\frac{v_i^{\ell} v_{M_I}^g v_{J_p}^{\ell}}{v_j^{\ell} v_{M_J}^g
v_{I_p}^{\ell}}.
\label{eq:modular spanning 2}
\end{equation}
If each module is homogeneous, we approximate $v_i^{\ell}\approx
v_{I_p}^{\ell}$, $v_{j}^{\ell}\approx v_{J_p}^{\ell}$ and obtain
$v_i/v_j\approx v_{M_I}^g/v_{M_J}^g$; the global structure of the 
network laid out by links across
modules determines the influence of each node. If each module is
heterogeneous in degree, we use the MA, \textit{i.e.},
$v_i^{\ell}\approx (k_i^{\rm out}/k_i^{\rm in})/ \sum_{i^{\prime};
\mbox{node }i^{\prime}\in M_I} (k_{i^{\prime}}^{\rm out}/ k_{i^{\prime}}^{\rm
in})$ and $v_j^{\ell}\approx (k_j^{\rm out}/k_j^{\rm in})/
\sum_{j^{\prime}; \mbox{node }j^{\prime}\in M_J} (k_{j^{\prime}}^{\rm out}/
k_{j^{\prime}}^{\rm in})$.  By assuming that $I_p$ ($J_p$) is a
typical node in $M_I$ ($M_J$), we set $v_{I_p}^{\ell}\approx
1/\sum_{i^{\prime}; \mbox{node }i^{\prime}\in M_I} 
(k_{i^{\prime}}^{\rm out}/
k_{i^{\prime}}^{\rm in})$ and $v_{J_p}^{\ell}\approx
1/\sum_{j^{\prime}; \mbox{node }j^{\prime}\in M_J}
(k_{j^{\prime}}^{\rm out}/
k_{j^{\prime}}^{\rm in})$.  Then, \EQ\eqref{eq:modular spanning 2} is
transformed into
\begin{equation}
\frac{v_i}{v_j} \approx
\frac{\left(k_i^{\rm
out}/k_i^{\rm in}\right)v_{M_I}^g}
{\left(k_j^{\rm
out}/k_j^{\rm in}\right)v_{M_J}^g}.
\label{eq:modular spanning 3}
\end{equation}
Therefore, we define an approximation scheme,
called the MA-Mod, for node $i$ in module $M_I$ as
\begin{equation}
v_i\propto \frac{k_i^{\rm out}}{k_i^{\rm in}}v_{M_I}^g.
\label{eq:modular spanning 4}
\end{equation}
Equation~\eqref{eq:modular spanning 4} can be used
for general modular networks
in which 
different modules can be connected by more than one links.

Two crucial assumptions underlie \EQ\eqref{eq:modular spanning 4}.
Firstly, a module is assumed to be
an uncorrelated and possibly heterogeneous random network
so that the MA is effective within the module. 
Note that the degree of nodes can be heterogeneously distributed.
Secondly, most links are assumed to be
intramodular so that the local MA
is simply given by $v_i^{\ell}\propto k_i^{\rm out}/k_i^{\rm in}$.

To obtain $v_{M_I}^g$ for general networks,
we define
$w_{I\to J}=
\sum_{i\in M_I, j\in M_J} w_{ij}$ and approximate
$v_i\approx v_{M_I}^g/\left[\sum_{I^{\prime}=1}^m N_{I^{\prime}}
v_{M_{I^{\prime}}}^g\right]$ (node $i\in M_I$), where $N_{I^{\prime}}$ is
the number of nodes in $M_{I^{\prime}}$. Then,
$\sum_{i=1}^N v_i=1$ is
satisfied.
Equation~\eqref{eq:influence} is transformed into
\begin{equation}
v_{M_I}^g = \frac{\sum_{J\neq I} w_{I\to J}v_{M_J}^g}{k_I^{\rm in}},
\quad (1\le I\le m),
\label{eq:influence for modules}
\end{equation}
where
\begin{equation}
k_I^{\rm in}\equiv\sum_{J^{\prime}\neq I} w_{J^{\prime}\to I}
= \sum_{i\in M_I, j\notin M_I} w_{ji}.
\end{equation}
Equation~\eqref{eq:influence for modules} 
has the same form as \EQ\eqref{eq:influence}.
By solving the set of $m$ linear equations,
we obtain $v_{M_I}^g$.

Equation~\eqref{eq:modular spanning 4} adapted for 
$v_i \left(G^{\rm rev}\right)$ leads to
$v_i\left(G^{\rm rev}\right)\approx
(k_i^{\rm in}/k_i^{\rm out}) v_{M_I}^g\left(G^{\rm rev}\right)$.
By combining this with \EQ\eqref{eq:R vs v}, we obtain
the MA-Mod scheme for the PageRank with $q=0$:
\begin{equation}
R_i\propto \frac{k_i^{\rm in}}{k_I^{\rm out}}R_{M_I}^g,
\label{eq:MA-Mod PR 2}
\end{equation}
where
\begin{equation}
k_I^{\rm out}\equiv \sum_{J^{\prime}\neq I} w_{I\to
J^{\prime}}= \sum_{i\in M_I, j\notin M_I} w_{ij}.
\end{equation}

\section{Application to Real Data}\label{sec:data}

We examine the effectiveness of
the MA-Mod scheme using three datasets from different fields.

\subsection{Neural network}

In the network of nematode \textit{C.~elegans}, a pair of neurons may
be connected by chemical synapses, which are directed links, or gap
junctions, which are undirected links. 
We calculate the influence of neurons on the basis of
a connectivity dataset \cite{Chen06pnas,wormatlas}.
The link weight $w_{ij}$ is assumed to be the sum of the number of
chemical synapses from neuron $i$ to neuron $j$ and that of
the gap junctions
between neuron $i$ and neuron $j$.  The following results are qualitatively
the same if we ignore the gap junction or the link weight (see Appendix~C
for the results).  The largest strongly connected component, which we
simply call the neural network, contains $274$ nodes and $2959$ links.

It is difficult to determine whether the influence or the PageRank 
is more appropriate from current biological
evidence. If postsynaptic
neurons linearly integrate different synaptic inputs,
%
%
the influence may be an
appropriate measure.  In contrast, postsynaptic neurons may
effectively select one synaptic input by a nonlinear mechanism.
If each input
is selected with the same probability in a long
run and the activity level does not differ much across neurons,
the PageRank may be appropriate. We examine both scenarios
using power iteration (see Appendix~D for the methodology).

Among 274 neurons, 54, 79, and 87 neurons are classified as sensory
neurons, interneurons, and motor neurons, respectively
\cite{wormatlas}.
By definition, sensory neurons directly receive
external input such as touch and chemical substances, motor neurons
send direct commands to move the body, and interneurons mediate
information processing in various ways. The other neurons are
polymodal neurons or neurons whose functions are unknown.  Neurons
with a large $v_i$ are mostly sensory neurons.  For example, among the
10 neurons with the largest $v_i$, 8 are sensory neurons (ALMR,
ASJL, ASJR, AVM, IL2VL, PHAL, PHAR, PVM) and 2 are interneurons
(AIML, AIMR).  Generally speaking, these neurons have a large $v_i$ not
simply because their $k_i^{\rm out}/k_i^{\rm in}$ is large.  The
average of $v_i/[(k_i^{\rm out}/k_i^{\rm in})/\sum_{j=1}^N(k_j^{\rm out}/k_j^{\rm in})]$ over the 10 neurons is
equal to 3.456 (see \TAB\ref{tab:cele best SI} in Appendix~C
for the values for individual neurons).
These neurons are located at upper levels of the
neural network in the global sense.
The conclusion remains qualitatively the same if we
use $R_i(G^{\rm rev})$. Recall that the PageRank is calculated
for $G^{\rm rev}$ because the meaning of the direction of the
link in the influence is opposite to that in the PageRank.

\begin{figure}
\begin{center}
\includegraphics[width=8cm]{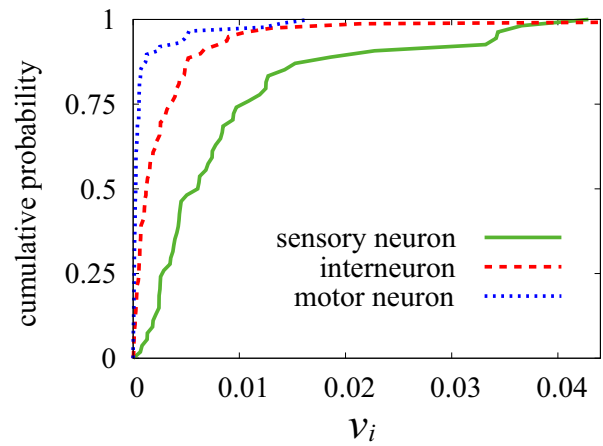}
\caption{Cumulative distribution of $v_i$ 
for 54 sensory neurons,
79 interneurons, and 87 motor neurons in
\textit{C.~elegans} neural network.}
\label{fig:cele histogram}
\end{center}
\end{figure}

The average values of $v_i$ ($R_i(G^{\rm rev})$) for sensory neurons,
interneurons, and motor neurons are equal to
0.009235 (0.006621), 0.003614 (0.005415), 
and 0.001032 (0.001323), respectively.
%
The cumulative distributions of $v_i$ for different classes of
neurons are shown in
\FIG\ref{fig:cele histogram}. 
Even though many
synapses from motor neurons to interneurons and sensory neurons,
and synapses from interneurons to sensory neurons exist,
these numerical results indicate that
the neural network is
principally hierarchical.
Generally speaking, sensory neurons,
which directly receive external stimuli,
are located at upper levels of the
hierarchy, motor neurons are
located at lower levels, and interneurons are located in between.
Sensory neurons serve as a source of signals flowing
to interneurons and motor neurons down the hierarchy.

\begin{figure}
\begin{center}
\includegraphics[width=8cm]{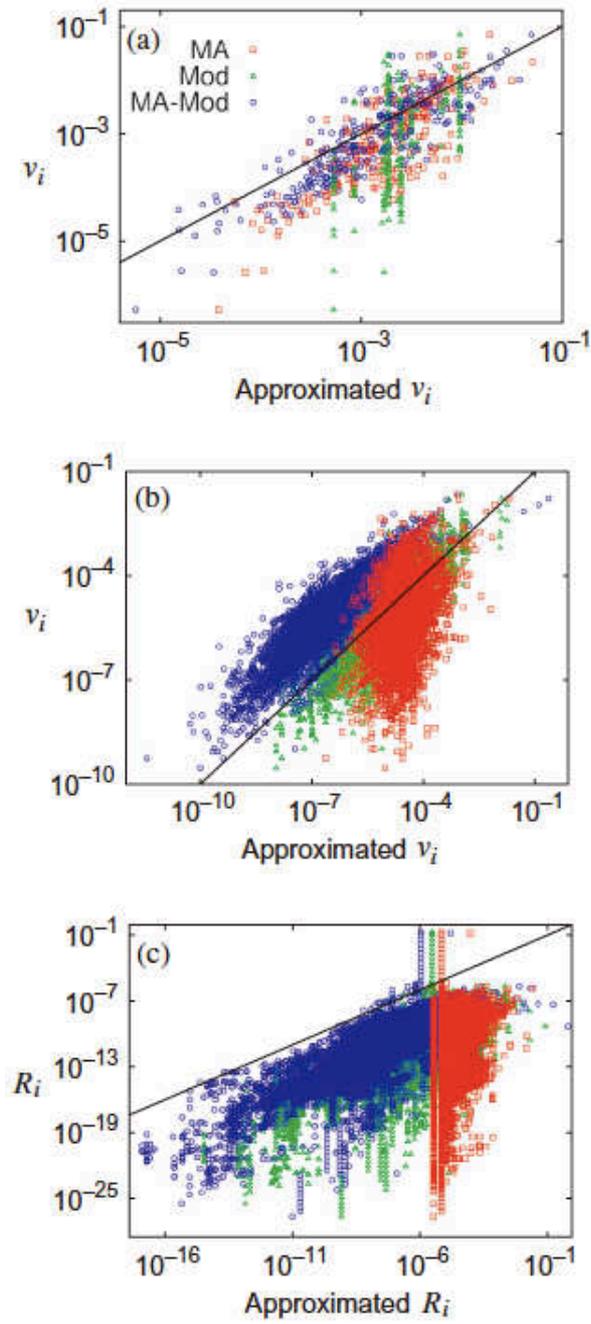}
\caption{(a) $v_i$ for neural network, (b) $v_i$ for email social
network, and (c) $R_i$ for WWW with $q=0$.  The quantities placed 
on the
horizontal axis are the MA (\textit{i.e.}, the normalized $k_i^{\rm
out}/k_i^{\rm in}$ for $v_i$ and the normalized $k_i^{\rm in}$ for
$R_i$) (red squares), Mod (green triangles), and MA-Mod (blue
circles).}
\label{fig:data}
\end{center}
\end{figure}

The relation between $v_i$ and the MA is shown in
\FIG\ref{fig:data}(a) by the squares. They appear 
strongly correlated.  
However,
the Pearson correlation
coefficient (PCC; see Appendix~E for definition) between
$v_i$ and the MA 
is not large ($=0.5389$), as shown in \TAB\ref{tab:modular PCC},
because $v_i$ tends to be larger than
the MA for nodes with a large $v_i$. Note that the 
data are plotted in the log-log scale in
\FIG\ref{fig:data}.

The neural network has modular structure \cite{Muller08}.
%
%
To use the MA-Mod scheme 
(\EQ\eqref{eq:modular spanning 4}), we apply a
community detection algorithm \cite{Rosvall08pnas} to the neural
network.
We have selected this algorithm \cite{Rosvall08pnas} because a
directed link in the present context indicates the flow rather than
the connectedness on which a recent algorithm
\cite{Leicht08prl} is based.
As a result, we obtain $m=13$ modules,
calculate $v_{M_1}^g$, $\ldots$, $v_{M_m}^g$ from the network of the
$m$ modules, and use \EQ\eqref{eq:modular spanning 4}.  $v_i$ is
plotted against the MA-Mod in \FIG\ref{fig:data}(a), indicated
by circles.
The data fitting has improved compared to the case of the MA,
in particular for small values of $v_i$.  The
PCC between $v_i$ and the MA-Mod is larger than that between
$v_i$ and the MA
(\TAB\ref{tab:modular PCC}). In this example, this holds true for the
raw data
and the logarithmic values of the raw data.
As a benchmark, we assess the performance of the global estimator
$v_i\approx v_{M_I}^g/\left[\sum_{I^{\prime}=1}^m N_{I^{\prime}}
v_{M_{I^{\prime}}}^g\right]$ (node $i\in M_I$), which we call the Mod.
The Mod ignores the variability of $v_i$ within the module and is
exact for networks with completely homogeneous modules, such as the
network shown in \FIG\ref{fig:example hie}.  The performance of the Mod
is poor in the neural network, as indicated by the
triangles in \FIG\ref{fig:data}(a) and the PCC listed in
\TAB\ref{tab:modular PCC}.

The values of the PCC between the actual and approximated 
$R_i(G^{\rm
rev})$ are also listed in \TAB\ref{tab:modular PCC}. The results
for the PageRank are qualitatively the same as those for the influence.
With both measures,
the module membership is a crucial determinant of
centralities of individual nodes. Note that, on the basis of the Mod
for the influence given by
\begin{equation}
v_i\approx
\frac{v_{M_I}^g}{\sum_{I^{\prime}=1}^m N_{I^{\prime}}
v_{M_{I^{\prime}}}^g},
\end{equation}
the Mod for the PageRank is given by
\begin{equation}
\frac{R_i(G^{\rm rev})}{k_i^{\rm in}}\approx 
\frac{R_{M_I}^g(G^{\rm
rev})}{k_I^{\rm in} \sum_{I^{\prime}=1}^m N_{I^{\prime}}
v_{M_{I^{\prime}}}^g},
\end{equation}
\textit{i.e.},
\begin{equation}
R_i(G^{\rm rev}) \propto
\frac{k_i^{\rm in}R_{M_I}^g(G^{\rm rev})}{k_I^{\rm in}}
\approx \frac{R_{M_I}^g(G^{\rm
rev})}{N_I}.
\end{equation}
We approximate $k_i^{\rm in}$ by $k_I^{\rm in}/N_I$
because the information about local degree is unavailable for the
Mod.

\begin{table}
\begin{center}
\caption{The Pearson correlation coefficient (PCC) between
the centrality measures and different estimators.}
\label{tab:modular PCC}
\begin{tabular}{|c|c|c|c|c|c|c|c|}\hline
network & \multicolumn{2}{|c|}{\textit{C.~elegans}}&
\multicolumn{2}{|c|}{Email}&
\multicolumn{3}{|c|}{WWW}\\ \hline
$N$ & \multicolumn{2}{|c|}{274}& \multicolumn{2}{|c|}{9079}&
\multicolumn{3}{|c|}{53968}\\ \hline
$m$ & \multicolumn{2}{|c|}{13}& \multicolumn{2}{|c|}{637}&
\multicolumn{3}{|c|}{2977}\\ \hline
centrality & $v_i$ & $R_i(G^{\rm rev})$ & $v_i$ & $R_i(G^{\rm rev})$ 
& $v_i(G^{\rm rev})$ & $R_i$ & $R_i$\\ \hline
$q$ & N/A &0& N/A &0& N/A &0& 0.15\\ \hline
MA&          0.5389& 0.3593& 0.5066& 0.3997& 0.0073& 0.0007& 0.2162\\
Mod&         0.2927& 0.4346& 0.5010& 0.2452& 0.0003& -0.0003& 0.4104\\
MA-Mod &     0.7295& 0.5005& 0.5066& 0.2671& 0.0000& 0.0000& 0.3166\\
MA (log)&    0.8024& 0.7073& 0.3636& 0.5353& 0.3109& 0.1289& 0.4627\\
Mod (log)&   0.5195& 0.5503& 0.8075& 0.7147& 0.7800& 0.7147& 0.4098\\
MA-Mod (log)&0.8736& 0.8252& 0.8798& 0.9022& 0.7964& 0.7812& 0.6256\\
\hline
\end{tabular}
\end{center}
\end{table}

\subsection{Email social network}

Our second example is
the largest
strongly connected component of an email social network
\cite{Ebel02_email}.  A directed link exists between a sender and a
recipient of an email. The network has modular structure
\cite{Palla07newjp}. In the weighted network that we consider here,
the link weight is defined by the number of emails.
The following results do not qualitatively
change even if we neglect the link weight (see Appendix~F).
The largest strongly
connected component has 9079 nodes and 23808 links and is 
partitioned into 637 modules.

Whether the influence or the PageRank
is appropriate for ranking nodes depends
on the assumption about human behavior. If recipients spend the same
amount of time on each incoming email (\textit{i.e.}, the 
link of weight
unity),
$v_i$ is relevant. In contrast, 
recipients may have a fixed amount of time for dealing
with all incoming emails. Then,
a recipient may equally distribute
the total time available to each email depending on
the number of incoming emails. Under this
assumption, the PageRank is relevant.
We analyze both $v_i$ and $R_i(G^{\rm rev})$.

In \FIG\ref{fig:data}(b),
the values of $v_i$ are plotted against those obtained by
different estimators. On the log-log scale, 
the MA-Mod performs considerably better
than the MA. Remarkably, even the
Mod, in which nodes in the same module share an estimated centrality
value, performs better
than the MA.  This is a strong indication that the
structure of the coarse-grained network of modules
is a more important determinant of $v_i$ than the local
structure (\textit{i.e.}, degree) in this example.  The values of the
PCC summarized in \TAB\ref{tab:modular PCC} support our
claim.
The PCC for the MA-Mod and
the Mod is considerably
larger than that for the MA on the logarithmic scale,
which implies that
the MA-Mod is especially effective for nodes with small $v_i$.
The values of the PCC between $R_i(G^{\rm rev})$ and the different
estimators are listed in \TAB\ref{tab:modular PCC}. These results are
qualitatively the same as those for $v_i$.

\subsection{WWW}\label{sub:www}

Our last example is the largest strongly connected components of a WWW
dataset \cite{Albert99}. The original network contains 325729 nodes
and 1469680 links, and the largest strongly connected component
contains 53968 nodes and 296229 links.  The MA fits the PageRank (with
$q>0$) of some WWW data when nodes of the same degree are grouped
together \cite{Fortunato06pnas%
} but not other data \cite{Donato04}. Because of the modular structure
of the WWW \cite{Palla07newjp}, the MA-Mod is expected to perform
better than the MA.

In \FIG\ref{fig:data}(c), $R_i$ for $q=0$ is
plotted against the MA, Mod, and MA-Mod.
For nodes with small PageRanks,
the MA-Mod, and even the Mod,
are considerably better correlaed with $R_i$ than the MA is 
(note the use of the log-log scale in
\FIG\ref{fig:data}(c); also see \TAB\ref{tab:modular PCC}).  These
nodes are located at lower levels of hierarchy.
%
The results are qualitatively the same if we use the influence
(\TAB\ref{tab:modular PCC}). Note that
we reverse the links and calculate $v_i(G^{\rm rev})$ because a directed link
in the WWW indicates an impact of the target node on the source node.

The MA-Mod for the
PageRank can be extended to the case $q>0$.
From \EQ\eqref{eq:page}, the MA for the PageRank is given by
\begin{equation}
R_i\approx \frac{q}{N}+(1-q)\frac{k_i^{\rm in}}{\left<k\right>}\bar{R}
= \frac{q+(1-q)\frac{k_i^{\rm in}}{\left<k\right>}}{N},
\label{eq:PageRank MA}
\end{equation}
which implies that $k_i^{\rm in}$ in the MA for $q=0$ is replaced by
$q\left<k\right>+(1-q)k_i^{\rm in}$ for general $q$.  We define the
MA-Mod for $q>0$ by
\begin{equation}
R_i\propto
\frac{\left[q\left<k\right>+\left(1-q\right) k_i^{\rm
in}\right]R_{M_I}^g}
{q \sum_J k_J^{\rm
out}/m+\left(1-q\right)k_I^{\rm out}}.
\end{equation}
Note that this ansatz
is heuristic, whereas \EQ\eqref{eq:MA-Mod PR 2} used for $q=0$
has an analytical basis. The PCCs between the PageRank with $q=0.15$
and the three estimates are listed in \TAB\ref{tab:modular PCC}.  The
MA-Mod performs better than the MA. The advantage of the MA-Mod over
the MA is smaller for $q=0.15$ than for $q=0$ because a larger $q$
implies a heavier neglect of the network structure.
%

The definition of the PageRank given by \EQ\eqref{eq:page} is not
continuous with respect to the outdegree; the term $\delta_{k_i^{\rm
out},0}(1-q)R_i$ is present for $k_i^{\rm out}=0$ (\textit{i.e.},
dangling node) and absent for $k_i^{\rm out}>0$. Therefore, dangling
nodes can have large PageRanks.  To improve the MA-Mod for $q>0$, we
should separately treat dangling nodes and other nodes in the same
module.  We do not explore this point because this situation seems to
be specific to the working definition of the PageRank.

In practice, nodes with a small $R_i$ could be irrelevant to the
performance of a search engine, which outputs a list of websites with
the largest PageRanks.  However, nodes with small PageRanks
constitute the majority of a network when the PageRank follows a
power-law distribution. This is the case for the real WWW data, which
are scale-free networks \cite{Fortunato06pnas,Donato04
}. Our method is considerably better than the MA especially for nodes
with small PageRanks.

In general, 
the WWW is nested, with each level defined by webpages,
directories, hosts, and domains. At the host level,
for example, most links are directed toward nodes
within the same host \cite{Kamvar03
}.
Therefore, a host can be regarded as a module in the network. 
By calculating the importance of the host, called the BlockRank,
the PageRank can be efficiently computed
\cite{Kamvar03}.
%
%
In spirit, our $R_{M_I}^g$ is similar to
the BlockRank, although our $R_{M_I}^g$ is used for identifying
the hierarchical levels of networks and systematically
approximating $R_i$.

It should be noted that, in general, our approximation scheme
runs much faster than the direct calculation of $v_i$ or $R_i$ for
large networks. This is because the community
detection algorithm \cite{Rosvall08pnas}
is fast and the power iteration
used for calculating $v_i$ and $R_i$ converges faster for a smaller
network in most (but not all) cases.  In the WWW, which is
a large network, our
approximation scheme for the PageRank with $q=0$ ran more than 100
times faster than the direct calculation on our computer.

\section{Discussion and Conclusions}

We have shown that the hierarchical structure of directed modular
networks considerably affects ranking-type centrality measures of
individual nodes.
Using the information about connectivity among modules,
we have significantly
improved the estimation of centrality values. Our
theoretical development is based on the measure that we have proposed
(\textit{i.e.}, influence), but the
conclusions hold true for both the influence and the PageRank.
Our method can be implemented for variants of the PageRank including
the eigenfactor \cite{Davis08,Fersht09}
and the so-called
invariant method \cite{Palacios04,Pinski76} used for ranking 
academic journals.
%
%

The hierarchy discussed in this study is different from the nestedness
of networks. Many networks are hierarchical in the sense that they are
nested and have multiple scales \cite{Ravasz02,Guimera05nat,Sales07,Clauset08nat}.
A modular network is hierarchical in this sense, at least to a
limited extent; two hierarchical levels are defined by the scale of
the entire network and that of a single module. In contrast, we are
concerned with hierarchical relationships among modules defined by the
directionality of networks.  This concept of hierarchy has been
studied for, for example, food webs \cite{Garlaschelli03nat},
transcription networks \cite{Lagomarsino07pnas}, and social dynamics
\cite{Castellano09}, but its understanding based on networks is
relatively poor in spite of its intuitive appeal.  The influence and
the PageRank quantify the hierarchical position of individual nodes
and of modules.

In real networks, nodes and links are subjected to changes.
Such changes affect nodes near the perturbed nodes, but may not
significantly affect modules. In social networks,
large groups change slowly over time as compared to
small groups \cite{Palla07nat}.
In addition, in the absence of complete knowledge of networks,
modest understanding of networks at the level of the modular structure may
be adequate. Nodes in a module may also have a common function.
%
These are main reasons behind investigating the
modular structure of networks.  We have shown that the modular
structure is also important in the context of directed networks,
hierarchy, and ranking.
The definition of module is complex in the case of directed
networks as compared to undirected networks, and module detection
in directed networks is currently under investigations
%
(see \cite{Fortunato09} for a review).
We hope that our results aid the development of the concept of
modules and related algorithms in directed networks.

\section*{Acknowledgments}

We thank Jesper Jansson and Kei Yura for their valuable discussions.
N.M. acknowledges the support through
the Grants-in-Aid for Scientific Research
(Nos. 20760258 and 20540382) from MEXT, Japan.

\renewcommand{\thetable}{A\arabic{table}}
\setcounter{table}{0}
\setcounter{page}{1}

\section*{Appendix A: Influence is obtained from various dynamical models on networks}

\subsection*{Fixation probability of evolutionary dynamics}\label{sub:LD}
$v_i$ represents the probability that an `opinion' introduced at node $i$
spreads to the entire network.  We consider stochastic competitive
dynamics between two equally strong types of opinions
$A$ and $B$; each node takes either $A$ or $B$ at a given time.  In
the so-called link dynamics (LD) \cite{Antal06prl,Sood08pre}, which is
a network version of the standard voter model, 
one link $(i,j)\in E$ is selected
for reproduction with an equal probability in each time step.  Then,
the type at node $i$ replaces that at node $j$. This process is repeated
until $A$ or $B$ takes over the entire network.

$v_i$ coincides with the fixation
probability denoted by 
$F_i^{\rm LD}$, which is the probability that a new type
$A$ introduced at node $i$ in the network of the resident type $B$ nodes
takes over the entire network \cite{Masuda09njp}.
To calculate $F_i^{\rm LD}$, fix a network and
consider the initial configuration in which $A$ is located at
node $i$ and $B$ is located at the other $N-1$ nodes.
In the first time step, one of
the following events occurs.
With the
probability $w_{ij}/\sum_{i^{\prime},j^{\prime}} w_{i^{\prime}
j^{\prime}}$,
the link $(i,j)\in E$ is selected for reproduction.
Then, type $A$ is located at nodes $i$ and $j$. 
Let $F_{\{i,j\}}^{\rm LD}$ denote the fixation
probability of type $A$ for this new configuration.
With the
probability $w_{ji}/\sum_{i^{\prime},j^{\prime}} w_{i^{\prime}
j^{\prime}}$,
the link $(j, i)\in E$ is selected,
type $A$ becomes extinct, and the dynamics terminates.
With the remaining probability
$\sum_{i^{\prime}\neq i,j^{\prime}\neq i} w_{i^{\prime}j^{\prime}}
/\sum_{i^{\prime},j^{\prime}} w_{i^{\prime}j^{\prime}}$,
the configuration of types $A$ and $B$ on the network does not
change.  Therefore, we
obtain
\begin{equation}
F_i^{\rm LD}=\sum_j \frac{w_{ij}}{\sum_{i^{\prime},j^{\prime}}
w_{i^{\prime}j^{\prime}}}
F_{\{i,j\}}^{\rm LD} +
\frac{\sum_j w_{ji}}{\sum_{i^{\prime},j^{\prime}} 
w_{i^{\prime}j^{\prime}}}
\times 0 +
\frac{\sum_{i^{\prime}\neq i,j^{\prime}\neq i}
  w_{i^{\prime}j^{\prime}}}
{\sum_{i^{\prime},j^{\prime}} w_{i^{\prime}j^{\prime}}}
F_i^{\rm LD}.
\label{eq:Fi LD elementary}
\end{equation}
Because $F_{\{i,j\}}^{\rm LD}=F_i^{\rm LD}+F_j^{\rm LD}$
\cite{Masuda09njp},
\EQ\eqref{eq:Fi LD elementary} leads to
\EQ\eqref{eq:influence} with $v_i$ replaced by $F_i^{\rm LD}$.

\subsection*{Continuous-time simple random walk}

Consider a simple random walk on the network in continuous time.
In a small time interval $\Delta t$,
a walker at node $i$ is attracted to its neighbor $j$, where
$(j,i)\in E$, with the probability $\Delta t$. 
Note that the direction of the link is opposite to 
the convention
because the directed link in the present study
indicates the influence of the source node of the link 
on the target node of the link.
The master equation for the density of the random walker at
node $i$, denoted by $F_i^{\rm RW}$ ($1\le i\le N$), is
represented by
\begin{equation}
\frac{dF_i^{\rm RW}}{dt} = \sum_{j=1}^N w_{ij}F_j^{\rm RW}
- k_i^{\rm in} F_i^{\rm RW}.
\label{eq:RW}
\end{equation}
Because the network $G$ is strongly connected, $F_i^{\rm RW}$ converges to
the unique stationary density. By setting the LHS of \EQ\eqref{eq:RW}
to 0, we obtain $F_i^{\rm RW}=v_i$.

The simple random walk is closely associated with the fixation
problem.  The so-called dual process of the LD is the coalescing
random walk. In the coalescing random walk, each of the $N$ walkers
basically performs the continuous-time simple random walk on the network
with the direction of all links reversed.  Therefore, the random
walker can traverse from node $i$ to node $j$ when $(j, i)\in E$. If two
random walkers meet on a node, they coalesce into one walker.  There
is only one walker after sufficiently long time, and the duality
between the two stochastic processes guarantees $F_i^{\rm RW}=F_i^{\rm LD}$
\cite{Masuda09njp}.

\subsection*{Reproductive value}

In population ecology, 
the number of offsprings an individual
contributes to
is quantified as the reproductive value of the individual.
The reproductive value of node $i$ is defined by
\EQ\eqref{eq:influence} \cite{Taylor90,Taylor96}. In practice,
a node represents a class of individuals defined
by, for example, sex, age, or habitat.

\subsection*{DeGroot model in social dynamics}

The DeGroot model \cite{Degroot74,
Jacksonbook,Olfati07}
is a discrete-time model
that represents the propagation of
information or opinions in social systems.
The state of the individual at node $i$ is represented by
a real value $p_i(t)$; $p_i(t)$ parameterizes
the information that the individual at node $i$ has at time $t$.
The weight $w_{ij}$ is the probability that
the individual at node $j$ copies the opinion at node $i$ in the next time
step. The normalization is given by $\sum^N_{i=1}w_{ij}=1$.
The states of the $N$ nodes
evolve according to
\begin{equation}
p_i(t)=\sum^N_{j=1}w_{ji} p_j(t-1).
\label{eq:DeGroot dynamics}
\end{equation}
If the network is strongly connected and aperiodic,
a consensus is reached asymptotically, \textit{i.e.},
$p_1(\infty)=\ldots=p_N(\infty)$ \cite{Jacksonbook,Olfati07}. 

The extent to which the initial information at
node $i$ influences the limiting common information in the
continuous-time version of the DeGroot model is equal to $v_i$.
To show this, we start with the discrete-time dynamics given by
\EQ\eqref{eq:DeGroot dynamics}. Suppose that
$F_i^{\rm DG}$ ($1\le i\le N$) satisfies $p_1(\infty)=\ldots=p_N(\infty)=
\sum^N_{i=1}F_i^{\rm DG}p_i(0)$ for arbitrary $p_1(0),\ldots,
p_N(0)$. Because the configuration $\{p_1(0),\ldots,p_N(0)\}$
and the configuration $\{p_1(1),\ldots,p_N(1)\}$ starting with
$\{p_1(0),\ldots,p_N(0)\}$ end up with
the identical $p_1(\infty)=\ldots=p_N(\infty)$,
we obtain
\begin{equation}
\sum^N_{i=1} F_i^{\rm DG}p_i(0)=
\sum^N_{i=1} F_i^{\rm DG}\left(\sum_{j=1}^N w_{ji}p_j(0)\right).
\end{equation}
Since $p_1(0),\ldots, p_N(0)$ are arbitrary, we obtain
\begin{equation}
F_i^{\rm DG}=\sum_{j=1}^N w_{ij}F_j^{\rm DG},
\label{eq:Fi^DG final}
\end{equation}

Equation~\eqref{eq:Fi^DG final} is of the same form as
\EQ\eqref{eq:influence}. However, the condition
$\sum_{i=1}^N w_{ij}=1$ is imposed in \EQ\eqref{eq:Fi^DG final}
because the dynamics are defined in the discrete time.
The continuous-time counterpart of the DeGroot model
is defined in \cite{Olfati07} as follows:
\begin{equation}
\frac{dp_i(t)}{dt}=\sum_{j=1}^N w_{ji}\left(p_j\left(t\right)
-p_i\left(t\right)\right).
\end{equation}
If $p_1(\infty)=\ldots=p_N(\infty)=
\sum^N_{i=1}F_i^{\rm DG}p_i(0)$, we obtain
$\sum_{i=1}^N F_i^{\rm DG}dp_i(t)/dt=0$, which leads to
\begin{equation}
\sum_{i=1}^N\left(\sum_{j=1}^N F_j^{\rm DG}w_{ij}-F_i^{\rm DG}
\sum_{j=1}^N w_{ji}\right)p_i(0)=0
\end{equation}
for arbitrary $p_1(0),\ldots, p_N(0)$. Therefore, $F_i^{\rm DG}=v_i$.

\subsection*{Collective responses in coupled oscillator dynamics}

According to \cite{Kori08}, consider
$N$ coupled phase oscillators
obeying
\begin{equation}
  \dot{\phi}_i = \omega_i
  + \sum_{j=1}^N \Gamma_{ij}\left( \phi_i - \phi_j \right)
  + \sigma p_i(t), \qquad
  i = 1,\cdots,N,
\label{eq:phase osc}
\end{equation}
where $\phi_i\in [0,2\pi)$ is the phase of
the oscillator $i$,
$\omega_i$ is the intrinsic
frequency of the oscillator $i$, $\Gamma_{ij}$
is the effect of node $j$ on node $i$, and $p_i(t)$ is
the input at time $t$
applied to node $i$. We assume that
(i) in the absence of the input (\textit{i.e.},
$\sigma=0$),
the system is fully phase-locked, \textit{i.e.},
$\phi_i = \phi_i^0
+ \Omega t$ for all $i$ with some
constants $\phi_i^0$ and $\Omega$ and that (ii)
the input is small, \textit{i.e.},
$\sigma \ll 1$, so that the system is
always close to the phase-locked state.
Using the synchronization condition, \textit{i.e.},
$\omega_i+\sum_{j=1}^N\Gamma_{ij}\left(
\phi_i^0-\phi_j^0\right)=\Omega$ ($1\le i\le N$),
which is implied by assumption (i), we linearize
Eq.~(\ref{eq:phase osc}) as
\begin{equation}
\dot{\psi}_i = \sum_{j=1}^N L_{ij} \psi_j + \sigma p_i(t),
\label{eq:phase osc linear}
\end{equation}
where $\psi_i\equiv \phi_i - \phi_i^0-\Omega t$ is a
small perturbation in the phase, and $L$ is the Jacobian
matrix given by
$L_{ij}$ $=$ $\left[ \sum_{j^{\prime} \ne i}\, 
\Gamma_{ij^{\prime}}^{\prime}
\left( \phi_i^0 - \phi_{j^{\prime}}^0 \right) \right]\delta_{ij}$ $-
\Gamma_{ij}'\left( \phi_i^0 - \phi_j^0 \right) \left( 1 - \delta_{ij}
\right)$.
Note that the effective weight of the link from
node $j$ to node $i$ is given by
$w_{ji}=-\Gamma_{ij}'\left( \phi_i^0 - \phi_j^0 \right)$.
Because assumption (i) implies the stability of the phase-locked
state, the real parts of all the eigenvalues of $L$
are negative, except a zero
eigenvalue. We define
the collective phase by
$\Theta \equiv \sum_{i=1}^N v_i \phi_i$.
Combination of $\sum_{i=1}^N v_i L_{ij}=0$ ($1\le j\le N$),
which is derived from \EQ\eqref{eq:influence}, 
and \EQ\eqref{eq:phase osc linear} yields
\begin{equation}
  \dot{\Theta} = \sum_{i=1}^N v_i \dot \phi_i
= \Omega + \sigma \sum_{i=1}^N v_i p_i(t).
   \label{eq:dot Theta}
 \end{equation}

Assumption (ii) implies
that $\dot\phi_1\approx \ldots\approx \dot\phi_N\approx
\dot\Theta$. Therefore, \EQ\eqref{eq:dot Theta} describes the dynamical
behavior of each oscillator and that of the entire network.
The response of the collective behavior to
the input applied to node $i$ is weighted by $v_i$.

\section*{Appendix B: Relationship between the influence and the PageRank}

To determine the relationship between the influence and the PageRank,
we rewrite \EQ\eqref{eq:page} as
\begin{equation}
R_i
=\sum_{j=1}^N \left[\frac{q}{N}+(1-q)\frac{w_{ji}}{k_j^{\rm out}}
+(1-q)\delta_{i,j}\delta_{k_i^{\rm out},0} \right] R_j.
\label{eq:page 2}
\end{equation}
From the original network $G$, define a complete and asymmetrically
weighted network $G^{\prime}$ using the matrix of link weights
$w^{\prime}_{ij}= q/N+(1-q)w_{ji}/k_j^{\rm out}
+(1-q)\delta_{i,j}\delta_{k_i^{\rm out},0}$.  Because $\sum_{j=1}^N
w^{\prime}_{ji}=1$ ($1\le i\le N$), $R_i$ in $G$ is equal to $v_i$
in $G^{\prime}$, which we denote by $v_i\left(G^{\prime}\right)$ for
clarity. Because self loops do not affect the calculation of the
influence, we can replace $w^{\prime}_{ij}$ by
$q/N+(1-q)w_{ji}/k_j^{\rm out}$.

In particular, $R_i$ for $q=0$ is equal to $v_i
\left(G^{\prime}\right)$, where $G^{\prime}$ is defined by
$w^{\prime}_{ij}=w_{ji}/k_j^{\rm out}$. In this case,
the PageRank and the influence are connected by the simple
relationship given by \EQ\eqref{eq:R vs v}.

\section*{Appendix C: Detailed analysis of the \textit{C.~elegans} neural network}

The relative contribution of a chemical synapse and that of
a gap junction to
signal transduction in the \textit{C.~elegans} neural circuitry are
unknown. 
In the main text, we have assumed that the neural network is a
weighted network in which a chemical synapse has the
same link weight as a gap junction. 
Here we examine three other 
variants of \textit{C.~elegans} neural networks.
In these three neural
networks, we neglect the link weight and/or gap junctions. The
omission of the link weight reflects the possibility that the
intensity of the
communication between two neurons may saturate
as the number of synapses increases. The omission of gap
junctions reflects the possibility that gap junctions may not
contribute to signal processing as significantly as chemical
synapses. Note that the largest strongly connected component shrinks to
a network of 237 nodes with 1936 synapses by the omission of
gap junctions.

For the three neural networks, the values of the PCC between the
centralities of the nodes and the three approximators are listed in
\TAB\ref{tab:cele PCC SI}.  We have examined both $v_i$ and $R_i
(G^{\rm rev})$ with $q=0$. In general, the MA-Mod predicts $v_i$
and $R_i(G^{\rm rev})$ better
than the MA in the three networks.  The results
listed in \TAB\ref{tab:cele PCC SI} are consistent with those presented
in the main text.

For the four neural networks, including the one in the main text,
the 10 most influential neurons are listed in \TAB\ref{tab:cele best SI}.
This list of 10 neurons is largely consistent across different
definitions of neural network. For the majority of these neurons,
$v_i$ is larger than the value predicted from the MA.

\section*{Appendix D: Power iteration}

If we use a standard numerical method such as the
Gaussian elimination,
the computation time required for calculating $v_i$ and $R_i$ from
\EQS\eqref{eq:influence} and \eqref{eq:page},
respectively, is $O(N^3)$. For sparse networks,
carrying out power iteration (also called Jacobi iteration)
may be much faster. The convergence of this iteration is guaranteed,
as explained below for the influence.
The proof for the PageRank is almost the same.

We rewrite \EQ\eqref{eq:influence} as
\begin{equation}
v_i = \sum_{j=1}^N \frac{w_{ij}}{{\sum_{j^{\prime}=1}^N}
w_{j^{\prime} i}} v_j.
\label{eq:power}
\end{equation}
Equation~\eqref{eq:power} indicates that
$v_i$ is the $i$-th element of the right eigenvector of the matrix
$M\equiv\left(M_{ij}\right) =\left(w_{ij}/\sum_{j^{\prime}=1}^N
w_{j^{\prime}i}
\right)$ for the eigenvalue equal to unity.
Multiplying $M$ by the diagonal matrix $(\delta_{ij}/\sum_{i^{\prime}=1}^N
w_{i^{\prime}j})$ on the 
right and its inverse on the left does not alter the
spectrum of $M$. This operation yields a new matrix whose ($i,j$) element
is given by
$(w_{ij}/\sum_{i^{\prime}=1}^N w_{i^{\prime}j})$.
The spectral radius of the new matrix is at most
unity because its maximum row sum matrix norm 
\cite[p.295]{Hornbook} is equal to unity.
Consequently, the spectral radius of $M$ is equal to unity.

Consider the power iteration scheme in which the $(t+1)$-th estimate
of $v_i$ ($1\le i\le N$) is given by the RHS of
\EQ\eqref{eq:power} in which the $t$-th estimate of $v_i$ ($1\le i\le
N$) is substituted.  If the network is strongly connected and
aperiodic, the nonnegative matrix $M$ is primitive, \textit{i.e.}, the
eigenvalue of the largest modulus, which is equal to unity in the
present case, is unique \cite[p.516]{Hornbook}.  Then, the convergence
of power iteration to the correct $\left(v_1, \ldots,
v_N\right)$ is guaranteed \cite[p.523]{Hornbook}.  The
Perron-Frobenius theorem \cite{Hornbook%
} guarantees that the Perron vector
$\left(v_1,\ldots,v_N\right)$ is uniquely
determined and that $v_i>0$ ($1\le i\le N$).  The power iteration
converges quickly if the modulus of the second eigenvalue of $M$ is
considerably smaller than that of the largest eigenvalue, \textit{i.e.},
unity.

\section*{Appendix E: PCC}

The PCC between $v_i$
and an estimator
 $v_i^{\rm est}$, such as MA, Mod, and MA-Mod,
is defined by
\begin{equation}
\frac{\frac{1}{N}\sum_{i=1}^N\left(v_i v_i^{\rm est}- 1/N^2\right)}
{\sqrt{\frac{1}{N}\sum_{i=1}^N\left(v_i-1/N\right)^2}
\sqrt{\frac{1}{N}\sum_{i=1}^N\left(v_i^{\rm est}-1/N\right)^2}}.
\label{eq:PCC}
\end{equation}
Note that $\sum_{i=1}^N v_i/N=\sum_{i=1}^N v_i^{\rm est}/N=1/N$.

\section*{Appendix F: Results for unweighted email social network}

The values of 
the PCC between the two centrality measures and different estimators
for the unweighted email social network are listed in
\TAB\ref{tab:PCC email unweighted}. The results are qualitatively the same 
as those for the weighted network shown in the main text.

\begin{table}
\begin{center}
\caption{PCC between
centrality measures and different estimators for \textit{C.~elegans}
neural networks.}
\label{tab:cele PCC SI}
\begin{tabular}{|c|c|c|c|c|c|c|}\hline
gap junction & \multicolumn{2}{|c|}{yes}&
\multicolumn{2}{|c|}{no}&
\multicolumn{2}{|c|}{no}\\ \hline
link & \multicolumn{2}{|c|}{unweighted}&
\multicolumn{2}{|c|}{weighted}&
\multicolumn{2}{|c|}{unweighted}\\ \hline
$N$ & \multicolumn{2}{|c|}{274}& \multicolumn{2}{|c|}{237}&
\multicolumn{2}{|c|}{237}\\ \hline
$m$ & \multicolumn{2}{|c|}{7}& \multicolumn{2}{|c|}{20}&
\multicolumn{2}{|c|}{15}\\ \hline
centrality & $v_i$ & $R_i(G^{\rm rev})$ & $v_i$ & $R_i(G^{\rm rev})$ &
$v_i$ & $R_i(G^{\rm rev})$\\ \hline
MA&          0.7420& 0.6331 & 0.2145& 0.0875 & 0.5153 & 0.4240\\
Mod&         0.3727& 0.2542 & 0.1577& 0.1741 & 0.1583 & 0.2224\\
MA-Mod &     0.8235& 0.7401 & 0.3328& 0.1189 & 0.4949 & 0.4659\\
MA (log)&    0.8478& 0.7942 & 0.6899& 0.6152 & 0.7976 & 0.7726\\
Mod (log)&   0.5190& 0.1937 & 0.7018& 0.7338 & 0.6743 & 0.6519\\
MA-Mod (log)&0.8995& 0.8752 & 0.8475& 0.8137 & 0.8785 & 0.8586\\ \hline
\end{tabular}
\end{center}
\end{table}

\begin{table}
\begin{center}
\caption{Most influential neurons in \textit{C.~elegans} neural
networks. $v_i/$(MA) indicates $v_i$ divided by the value obtained
from the MA.}
\label{tab:cele best SI}
\hspace*{-1.5cm}
\begin{tabular}{|c|c|c|c|c|c|c|c|c|c|c|c|}\hline
\multicolumn{3}{|c|}{with gap junction}&
\multicolumn{3}{|c|}{with gap junction}&\multicolumn{3}{|c|}{no gap junction}
&\multicolumn{3}{|c|}{no gap junction}\\ \hline
\multicolumn{3}{|c|}{weighted}&\multicolumn{3}{|c|}{unweighted}
&\multicolumn{3}{|c|}{weighted}&\multicolumn{3}{|c|}{unweighted}\\ \hline
neuron & $v_i$ & $\frac{v_i}{\rm (MA)}$&
neuron & $v_i$ & $\frac{v_i}{\rm (MA)}$&
neuron & $v_i$ & $\frac{v_i}{\rm (MA)}$&
neuron & $v_i$ & $\frac{v_i}{\rm (MA)}$\\ \hline
AIMR& 0.08876& 4.160&
PHAL&  0.04279& 3.226& AIMR& 0.06841& 3.924&VC04&  0.05899& 4.594\\
ASJL& 0.04287& 3.588&
PHAR&  0.04117& 3.449& ASJL& 0.04835& 3.467&VC05&  0.04439& 3.841\\
ALMR& 0.03657& 3.296&
AIMR&  0.04062& 2.356& ALMR& 0.03965& 2.843&AIMR&  0.03718& 2.227\\
PHAR& 0.03435& 7.740&
ASIL&  0.02748& 2.072& VC04& 0.03334& 2.988&AIML&  0.02722& 1.325\\
PHAL& 0.03419& 6.259&
ASIR&  0.02695& 2.540& PVM & 0.03246& 2.116&AWAL&  0.02715& 1.510\\
ASJR& 0.03319& 4.094&
AIML&  0.02152& 1.432& AVM & 0.02847& 1.047&AVG&   0.02426& 2.951\\
IL2VL& 0.02647& 0.456&
IL2VL& 0.02061& 0.706& AIML& 0.02304& 2.447&AVM&   0.01715& 1.028\\
AVM& 0.02273& 1.816&
ALMR&  0.01982& 2.135& AVG & 0.02257& 8.826&ASKR&  0.01701& 3.975\\
AIML& 0.02133& 2.231&
VC05&  0.01719& 2.160& ASJR& 0.02217& 2.649&ALMR&  0.01692& 1.647\\
PVM& 0.01860& 1.816&
VC04&  0.01505& 2.838& ADLL& 0.01777& 0.593&IL2VL& 0.01546& 0.669\\ \hline
\end{tabular}
\end{center}
\end{table}

\begin{table}
\begin{center}
\caption{PCC between
centrality measures and different estimators for unweighted email
social network.}
\label{tab:PCC email unweighted}
\begin{tabular}{|c|c|c|}\hline
$N$ & \multicolumn{2}{|c|}{9079}\\ \hline
$m$ & \multicolumn{2}{|c|}{599}\\ \hline
centrality & $v_i$ & $R_i(G^{\rm rev})$\\ \hline
$q$ & N/A &0\\ \hline
MA&          0.6628& 0.5536\\
Mod&         0.6537& 0.3290\\
MA-Mod &     0.6692& 0.4774\\
MA (log)&    0.2552& 0.4203\\
Mod (log)&   0.8719& 0.7754\\
MA-Mod (log)&0.8898& 0.9042\\ \hline
\end{tabular}
\end{center}
\end{table}

\newpage
\clearpage

\clearpage

\clearpage

\clearpage

\end{document}